\documentclass[letterpaper]{article} % DO NOT CHANGE THIS
\usepackage{aaai24}  % DO NOT CHANGE THIS
\usepackage{times}  % DO NOT CHANGE THIS
\usepackage{helvet}  % DO NOT CHANGE THIS
\usepackage{courier}  % DO NOT CHANGE THIS
\usepackage[hyphens]{url}  % DO NOT CHANGE THIS
\usepackage{graphicx} % DO NOT CHANGE THIS
\usepackage{natbib}  % DO NOT CHANGE THIS AND DO NOT ADD ANY OPTIONS TO IT
\usepackage{caption} % DO NOT CHANGE THIS AND DO NOT ADD ANY OPTIONS TO IT
\usepackage{algorithm}
\usepackage{algorithmic}

\usepackage{newfloat}
\usepackage{listings}

\usepackage{booktabs} % For formal tables
\usepackage{caption}
\usepackage{adjustbox}
\usepackage{subcaption}
\usepackage{relsize}
\usepackage{float}
\usepackage{url}
\usepackage{epstopdf}
\usepackage{graphics,graphicx}
\usepackage{array}
\usepackage{pifont}
\usepackage{multirow}
\usepackage{svg}
\usepackage{amsmath}
\usepackage{footmisc}
\usepackage{paralist, tabularx}
\usepackage{xcolor}

\DeclareCaptionStyle{ruled}{labelfont=normalfont,labelsep=colon,strut=off} % DO NOT CHANGE THIS
\floatstyle{ruled}
\newfloat{listing}{tb}{lst}{}
\floatname{listing}{Listing}
\title{Sponsored is the New Organic: Implications of Sponsored Results\\ on Quality of Search Results in the Amazon Marketplace}

\author {
    Abhisek Dash\textsuperscript{\rm 1},
    Saptarshi Ghosh\textsuperscript{\rm 2},
    Animesh Mukherjee\textsuperscript{\rm 2},\\
    Abhijnan Chakraborty\textsuperscript{\rm 2},
    Krishna P. Gummadi\textsuperscript{\rm 1} 
}
\affiliations {
     \textsuperscript{\rm 1}Max Planck Institute for Software Systems, Germany\\
    \textsuperscript{\rm 2}Indian Institute of Technology Kharagpur, India\\
    adash@mpi-sws.org, saptarshi@cse.iitkgp.ac.in, animeshm@cse.iitkgp.ac.in,\\abhijnan@cse.iitkgp.ac.in, gummadi@mpi-sws.org
}

\begin{document}

\maketitle

\begin{abstract}
Interleaving sponsored results (advertisements) amongst organic results on search engine result pages (SERP) has become a common practice across multiple digital platforms. Advertisements have catered to consumer satisfaction and fostered competition in digital public spaces; making them an appealing gateway for businesses to reach their consumers. 
However, especially in the context of digital marketplaces, due to the competitive nature of the sponsored results with the organic ones, multiple unwanted repercussions have surfaced affecting different stakeholders. From the consumers' perspective the sponsored ads/results may cause degradation of search quality and nudge consumers to potentially irrelevant and costlier products. 
The sponsored ads may also affect the level playing field of the competition in the marketplaces among sellers. 
To understand and unravel these potential concerns, we analyse the Amazon digital marketplace in four different countries by simulating 4,800 search operations. 
Our analyses over SERPs consisting 2M organic and 638K sponsored results show items with poor organic ranks (beyond 100$^\textrm{th}$ position) appear as sponsored results even before the top organic results on the first page of Amazon SERP. 
Moreover, we also observe that in majority of the cases, these top sponsored results are costlier and are of poorer quality than the top organic results. We believe these observations can motivate researchers for further deliberation to bring in more transparency and guard rails in the advertising practices followed in digital marketplaces.
%\red{10 pages + additional page for reference followed by appendices. }
\end{abstract}

\section{Introduction}~\label{sec: Intro}
Digital marketplaces have emerged as ubiquitous platforms to serve the purchase needs of consumers and livelihood of businesses across the globe\footnote{In this paper, we shall discuss the different nuances by keeping Amazon marketplace as a case study. Most of the practices discussed here are also deployed by other e-commerce marketplaces.}.
These platforms mediate the interaction between consumers and businesses through sophisticated algorithmic systems and choice architectures~\cite{dash2021when, dash2022alexa}. 
In recent years, \textit{advertising} has evolved as a lucrative gateway for businesses to reach potential consumers on online marketplaces. For instance, advertisements on Amazon are said to reach almost 96\% of all Americans between the age of 25 and 54~\cite{FTC2023Federal}. 

\begin{figure*}[t]
	\centering
		%\includesvg[width=\textwidth]{figures/Amazon-SERP-BabyWipes.svg}
            \includegraphics[width=\textwidth]{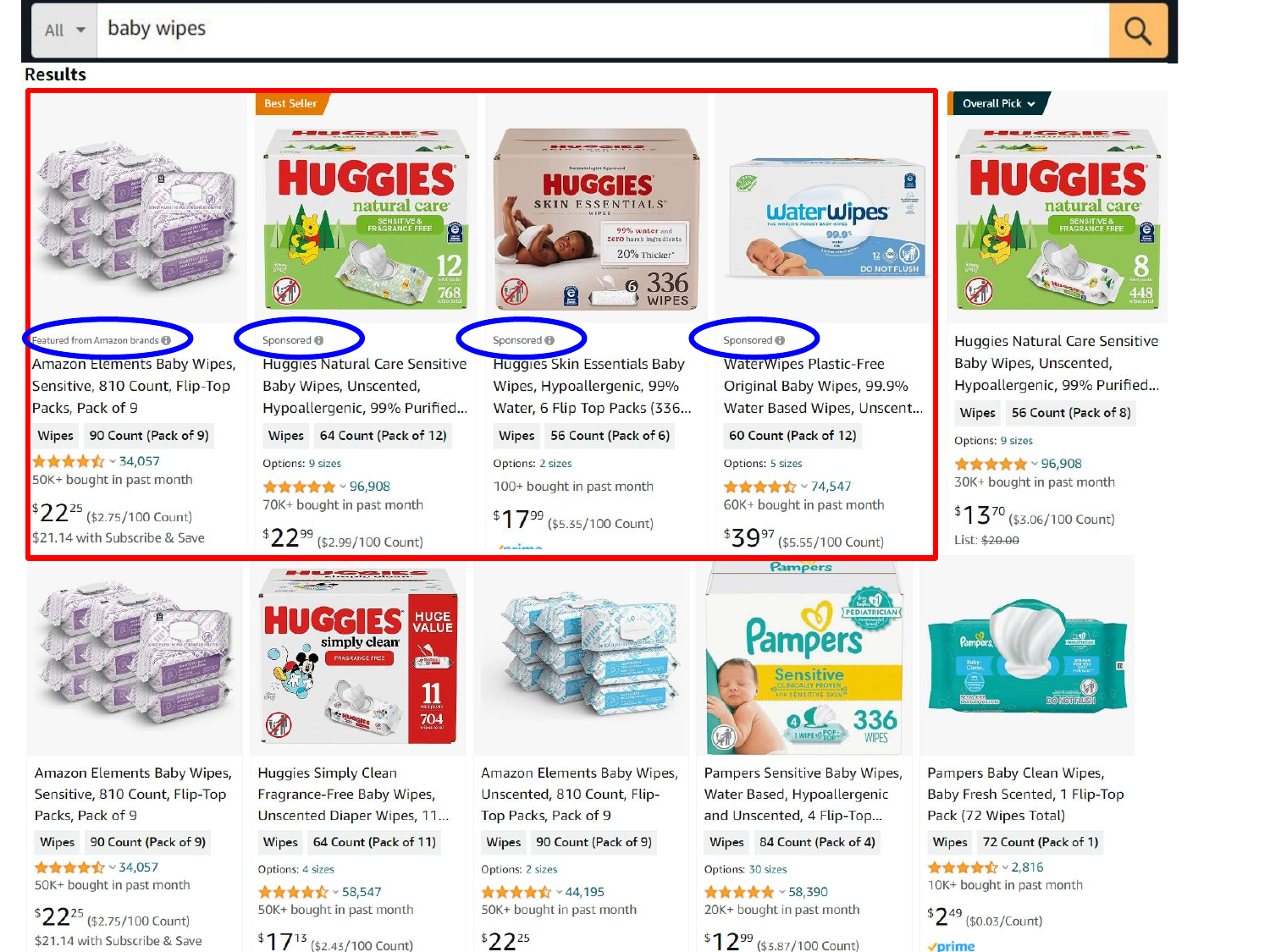}
         \caption{\textbf{First two rows of the search engine result page (SERP) for query `baby wipes' on {\tt Amazon.com}. The first four results are sponsored results (i.e., ads) while the rest of the results shown are organic.}}
    \label{Fig: SERP}
\end{figure*}

\noindent
\textbf{Sponsored product advertisements on Amazon: }
When a consumer searches for a query on Amazon, a number of results are shown in different arrangements on its search engine result pages (SERPs) in the decreasing order of relevance~\cite{sorokina2016amazon}. Increasingly, some of the results are marked with small gray tags that read `Sponsored'. These are sponsored product advertisements~\cite{Amazon2024Sponsored} on Amazon. These sponsored results can be looked as a way of providing exposure to sellers who are getting started on the marketplace and would find it harder to get better organic ranks~\cite{Amazon2024Sponsored}. In fact, many sellers have utilised sponsored product ads for improving their sales and revenue. Amazon also ensures that only those sellers' advertisements show up as sponsored results who are eligible to win the buy box of Amazon i.e., they are amongst the better sellers who as per Amazon's estimate can provide good quality of service to consumers~\cite{Amazon2024SP}.

However, increasingly sponsored results and their placements are gaining attention among policymakers. Over time, these sponsored results are replacing organic results at the top of SERPs. 
For example, Figure~\ref{Fig: SERP} shows the first two rows of search results for the query `baby wipes' on Amazon. 
The first four results (highlighted within red rectangle) on the SERP are all sponsored results i.e., advertisements. 
Each sponsored result is demarcated with a tag which has been encircled in blue. 
The organic results start at the right hand corner of the top row. 
Note that the organic results are ranked based on prior user engagement (clicks, product page visits, purchases, etc.) with different products while searching for the same query~\cite{sorokina2016amazon}. 
Further, the same product can appear as sponsored and organic results on the SERP as well. For example, the first product in Figure~\ref{Fig: SERP} also appears as a result in the SERP in organic position \textit{two}. 
In this paper, we consider that this first sponsored result has an organic rank \textit{two}. 
Next we discuss some potential concerns with sponsored ads keeping this example figure in the perspective.

\noindent
\textbf{Potential concerns from the point of view of consumers:} Note that apart from the first product (which costs $22.25\$$), none of the other sponsored results are among the top-5 organic results shown on the SERP. Due to cognitive biases, most consumers tend to give more attention to the top left of a web-page~\cite{baeza2018bias}. Similarly, multiple works have shown results appearing higher in the ranked lists are perceived to be more relevant and are more likely to be clicked~\cite{biega2018equity, singh2018fairness}. 
Hence, by being positioned before the organic results, sponsored ads are more likely to be clicked (or even purchased) by consumers. Although, in Amazon's defense, these results are demarcated as `sponsored', prior surveys have shown that almost half of the consumers on Amazon are not able to distinguish between sponsored and organic results~\cite{Graham2019Amazon}. On most occasions, these demarcations are in significantly smaller fonts and in gray color (as seen in Figure~\ref{Fig: SERP}). 
Even worse, sometimes it may not even read `Sponsored' but as `Featured from Amazon brands', as in the first result in Figure~\ref{Fig: SERP}. 
Replacing top organic results with relatively lower quality sponsored results (without proper demarcation) can potentially nudge consumers to irrelevant products and lead to consumer dissatisfaction. In fact, Amazon's internal surveys also hint at inclusion of sponsored results affecting consumer satisfaction. However, the ad revenue outweighs the loss that Amazon has endured due to degradation of search result quality~\cite{FTC2023Federal}; giving further incentive to Amazon to increase the ad-estate on its SERPs.

Now, let us consider the price of the different sponsored results and organic results in Figure~\ref{Fig: SERP}. The average price of the four sponsored results is $25.8\$$; whereas that of the same number of top organic results is $18.83\$$. Hence, the average price of sponsored results is 37\% costlier than the average price of organic results. 
Although this is an anecdotal example, these concerns are also raised in the lawsuit filed by FTC against Amazon late last year~\cite{FTC2023Federal}. If this phenomenon is a frequent occurrence, then sponsored ads may systemically nudge consumers towards costlier products.

\noindent
\textbf{Potential concerns from the point of view of sellers: }
Replacing and / or interleaving sponsored results among organic results demotes products from their organic and deserved position that they merit. For example, the Pamper sensitive baby wipes (shown as the penultimate result in Figure~\ref{Fig: SERP}) would be placed in the first row had there been no sponsored results at all. Consumers would not require to scroll down to get to see the product on most standard laptop or desktop devices. Such deprivation of exposure may have effects on the sales and revenue of the corresponding sellers. 

Furthermore, sponsored results being the first point of interaction for consumers, sellers and businesses will have more incentives to use advertisements as their gateway to reach consumers. If so, this will end up favoring the \textit{larger sellers} and putting the new and small scale sellers at disadvantageous position. 
Since all these ads are allocated based on second price auctions~\cite{Amazon2024Sponsored}, the seller bidding the highest price will eventually end up getting more exposure. 
This, in turn, (a)~may result in affecting the competition in the marketplace and (b)~the increased cost on advertising may trickle down to consumers in terms of increased prices.

\noindent
\textbf{Concerns resonated by policymakers:} The concerns described above have been resonated by policymakers across the globe. In the past few years, Amazon has faced multiple antitrust investigations~\cite{Amazon2019Online, Amazon2020Questions, EU2020Antitrust, FTC2023Federal}. Most of the antitrust investigations focus on Amazon's preferential treatment toward itself or some related sellers. 
However, we fear that some of the practices described above can have much more fundamental ramifications in the functioning of a fair marketplace. 
Even without the existence of any special relationships, interleaving and replacing organic results with potentially less relevant and costlier ads can cause harm to consumers and competition on the digital marketplaces.

\noindent
\textbf{Current work: }
In this work, we carry out a comprehensive study of the \textit{effects and implications (if any) of sponsored results on the quality of search results in digital marketplaces}. As a case study, we conduct our investigations on four different marketplaces of Amazon i.e., in {\tt Amazon.com} (USA), {\tt Amazon.in} (India), {\tt Amazon.de} (Germany) and {\tt Amazon.fr} (France). By designing web-scrapers, we collect search result data from each of these marketplaces. 
Our extensive analyses on 4,800 search operations and the resultant SERPs consisting of more than 2M organic results and 638K sponsored results bring out multiple concerning observations regarding top Amazon search results:

\begin{itemize}

    \item Nearly 30\% of results in our collected dataset on Amazon are sponsored results i.e., advertisements. In 85\% cases at least one sponsored result appears \textit{before} the top organic result on the first SERP.
    
    \item Many of the top sponsored results appearing in the observed SERPs seem to be less relevant than the organic results as per Amazon's own ranking algorithm. Almost half of the sponsored results never appear within the first three pages of the collected organic results.

    \item In more than half of the instances, the top sponsored results are 50\% costlier than the corresponding top organic results in our dataset. In other words, through positioning sponsored results before organic results, consumers may potentially be nudged to seemingly costlier products.

\end{itemize}

\noindent
\textbf{Curious case of ads on digital marketplaces:} We want to clarify that we do not posit that advertising in all cases is harmful to consumers and fair competition. In fact, if properly conducted, advertising can improve consumer satisfaction and foster healthy competition on the marketplaces. 
However, the case of digital marketplaces is a tricky one. 
If we notice Figure~\ref{Fig: SERP} again, the same product is appearing as the first sponsored result and the second organic result, i.e., the candidates of sponsored results and organic results are not mutually exclusive or complementary. 
In fact, in most cases they are competing offers or products. 
In other words, advertisements in digital marketplaces are \textit{competitive} in nature with the organic results~\cite{dash2021when}. 

Further, there can be significant delayed impact of these sponsored result on organic ranking as well. For example, as per Amazon's internal experiments, advertised products are 46 times more likely to be clicked on as opposed to non-advertised products~\cite{FTC2023Federal}. At the same time, click-through and several other browsing patterns are important metrics in deciding organic ranking~\cite{sorokina2016amazon}. Thus, if organic results on top of the page are increasingly replaced by sponsored results, it may not only propel some of the short term harms to consumers and businesses at scale, but also result in long term harms to the organic competition in the marketplace. Similar concerns were also raised by \citet{page1999pagerank} in their seminal work on PageRank algorithm, where they discussed how commercial interests can be a threat to organic ranking and the primary goals of a search system.
In fact, our observations in this paper demonstrate that Amazon's practices in showing sponsored ads in their SERPs may not arguably comply with several regulations on fair commercial practices in many countries. We discuss these legal aspects in the penultimate section of the paper.
Hence, we posit such fairness concerns should not go unnoticed by auditors, policymakers, practitioners and the platform companies as well.
\section{Related work}~\label{sec: Related}
In this section, we survey some prior works especially in the area of (a)~auditing digital marketplaces, (b)~algorithmic auditing of content search, and (c)~regulations on fair commercial practices to further contextualise the current work.

\subsection{Auditing digital marketplaces}
While Algorithmic auditing has emerged as a vibrant new direction of research under the broad umbrella of \textit{Responsible AI}, most of the prior works have come in the context of digital public space~\cite{kulshrestha2017quantifying, ali2019discrimination} and sensitive applications e.g., recidivism prediction~\cite{angwin2022machine}; gender detection~\cite{buolamwini2018gender, jaiswal2022two}; algorithmic hiring~\cite{wilson2021building} etc. Thus a long line of research has focused on fairness concerns regarding different sensitive demographics. However, the context of \textit{digital marketplaces} has been largely understudied. Recently, some studies have tried to understand the intricacies and fairness concerns in digital marketplaces~\cite{dash2024antitrust} focusing on their (a)~related item recommendations~\cite{dash2021when}, (b)~choice architectures~\cite{dash2024antitrust}. These works delve into the anti-competitive concerns in the digital marketplace. Whereas, several other studies look into issues like algorithmic pricing~\cite{chen2016empirical}, fairness across different modalities~\cite{dash2022alexa}, vaccine misinformation~\cite{juneja2021auditing} on Amazon.

In contrast, the current study focuses on the most important apparatus on Amazon ecosystem i.e., Amazon search with an aim to understand how sponsored advertising may potentially distort the observed ranking of products and what are their repercussions to the quality of search results.

\subsection{Algorithmic auditing of content search}
Algorithmic auditing is not new in the context of content search. Multiple prior audit studies have been conducted with different goals in mind. \citet{mehrotra2017auditing} studied whether Bing search engine provides equal level of satisfaction to its end-users. \citet{kulshrestha2017quantifying} proposed a bias quantification metric to evaluate the political biases. 
The same metric was then utilised to evaluate the partisanship of Google search result as well~\cite{robertson2018auditing}. While bias in search results has been a well-studied problem, interleaving of sponsored results and their effects has not yet been well studied. The interleaving of sponsored results become even more important in the context of digital marketplaces such as Amazon where the advertisements are traditionally lower funnel and are competitive in nature with the organic results~\cite{dash2021when}. Hence the effects of such interleaving should not go unnoticed.

\subsection{Regulations on fair commercial practices}
Maintaining fairness in marketplaces (both physical and digital) is of paramount importance. Thus a series of such regulations are in place in different parts of the world. Older regulations e.g., the Federal Trade Commission Act~\cite{FTC2006Federal} in the USA; Unfair commercial directives~\cite{EU2021Unfair} in the EU; Consumer Protection Act~\cite{DCA2019Consumer} in India -- all prohibit adopting unfair and deceptive commercial practices in the marketplaces. 
However, with increasing reliance on digital marketplaces, many countries are coming up with new directives to set guidelines for fair conduct on the online world to avoid deceptive user interface design and bring in transparency in deployed design choices and algorithms e.g., Dark Pattern directives in India~\cite{DCA2019Consumer} and DSA~\cite{EU2022DSA},  DMA~\cite{EC2022DMA} in the EU. 
We shall discuss some specific regulations in detail later in the paper, when we outline the legal implications of our observations.
%\vspace{-12 mm}
\section{Data collection}  \label{sec:dataset}

%\textbf{Platforms from which data is collected:}
Amazon search system is context sensitive~\citep{sorokina2016amazon} i.e., it uses different search ranking model for different contexts, where 
contexts are defined by the category of products and the specific Amazon marketplaces (e.g., USA or India etc.).
To this end, our data collection strategy covers more than ten product categories across four popular Amazon marketplaces across the globe -- {\tt Amazon.in} (India), {\tt Amazon.com} (USA), {\tt Amazon.de} (Germany) and {\tt Amazon.fr} (France). The increasing attention of policymakers toward digital marketplaces in these countries has motivated us to study these marketplaces.

\vspace{1mm}
\noindent 
\textbf{Search queries in two settings - default and refined:}
We start our data collection by searching for several queries on Amazon's desktop search across the four marketplaces. 
We observe that for all queries, irrespective of the query category, we get a maximum of 306 organic results. Amazon search box has several refinement settings as per the different product categories available on the marketplace. The default search setting is `All categories' which is used by most end-users.  
In this setting, only 306 organic results are shown for a given query, i.e., it does \textit{not} show an exhaustive list of products for the query.

However, for the purpose of our analyses, we need to know the organic position of as many sponsored products as possible; even beyond 306.
Hence, we choose to collect data in the \textit{refined setting} where we specify a category for the search query. Through this setting, one can have access to more products relevant to the query under the specified category. On the website of each of the four Amazon marketplaces stated earlier, we search for a set of 50 queries using the refined setting. The selected queries are among the most searched queries on Amazon~\cite{Hardwick2021Top} and also belong to some of the top and most competitive product categories on Amazon~\cite{Connolly2024Category}. The number of queries across the different top categories are shown in the upper panel of Table~\ref{Tab: QueryStat}. In this setting, we collect results for 50 queries across 4 different marketplaces at \textit{4 different temporal points} (which we refer to as \textit{snapshots}).

However, due to the overarching nature of the default setting, most end-users may actually query in the default setting and hence are exposed to search result pages returned in the default setting. Hence, we also collect one snapshot of search results for 1000 queries in the default setting, to investigate if observations in those SERPs conform to what we observe in the refined setting. The set of 1000 queries has been inherited from a prior work~\citet{dash2022alexa}. This set of queries cover some of the most competitive and popular categories e.g., electronics, computer accessories, home and kitchen, etc. The distribution of the queries across categories are shown in the lower panel of Table~\ref{Tab: QueryStat}.

\begin{table}[t]
	\noindent
	%\small
	\scriptsize
	\centering
	\begin{tabular}{ |p{2.6 cm}|r||p{2.5 cm}|r|}
		\hline
		{\bf Category} & \# Query & {\bf Category} & \# Query\\
            \hline \hline
            \multicolumn{4}{|c|}{Refined setting (50 queries $\times$ 4 snapshots $\times$ 4 marketplaces)}\\
		\hline
		Electronics & 9 & 	Mobile accessories & 3\\
		\hline
		Computer accessories & 8 & Office products & 3\\
		\hline
		Home \& Kitchen & 9 & Home improvement & 2\\
		\hline
		Health \& personal care & 8 & Video games & 2\\
		\hline
		Amazon devices & 4 & Others & 2 \\
            \hline \hline
            \multicolumn{4}{|c|}{Default setting (1000 queries $\times$ 1 snapshot $\times$ 4 marketplaces)}\\
		\hline
            Electronics & 238 & 	Mobile accessories & 74\\
		\hline
		Computer accessories & 211 & Office products & 48\\
		\hline
		Home \& Kitchen & 118 & Home improvement & 50\\
		\hline
		Health \& personal care & 112 & Video games & 25\\
		\hline
		Sports, fitness & 21 & Luggage and bags & 79 \\
		\hline
		%\hline
	\end{tabular}	
	\caption{{\bf Distribution of queries from top-10 categories present in our dataset. For these 50 queries, we collected data for 4 countries across 4 different timestamps. For a super set of 1000 queries, we also collect data for one snapshot across the 4 countries.}}
    \label{Tab: QueryStat}
	%\vspace{-8 mm}
\end{table}

\vspace{1mm}
\noindent \textbf{Crawling methodology:} To this end, we design a web-scraper to mimic search operations on Amazon desktop websites. The scraper is seeded with a query, and the corresponding search results are scraped. We perform this task till we get the last indexed SERP on Amazon for the given query. This enables us to find the organic search rank (and thus the \textit{Amazon defined relevance}) of each product that Amazon indexes at a certain point of time with respect to the mentioned query. We use Gecko driver and Mozilla Firefox browser for our data collection. 

We take some precautions to ensure the meaningfulness of the comparative analysis.  
The search algorithms are likely to tailor results based on several signals, including the geographical location, consumer browsing history, and so on. To minimize these variations, we collect search results over four different weeks through incognito Firefox browsers, during the same time frame, with the same delivery address and keeping all the necessary contexts e.g, account, IP address, delivery address etc. as similar as possible. 

\vspace{1mm}
\noindent \textbf{Data collected:} Upon searching a query on Amazon, a list of products are shown on the search engine result pages (SERPs). The search results are organised in different pages with several number of results per page. We collect the search results for each of the queries till we find no next available SERP. 
As stated earlier, Amazon shows two types of search results -- organic (algorithmically curated results considering prior consumer behavior) and sponsored ads. For each result, we collect its type. 
For each product shown as a search result, we also collect different metadata, e.g., brand, price, average user rating, number of reviews, etc.

\vspace{1mm}
\noindent Next, we will investigate how sponsored search results may affect the quality of Amazon search results.

\section{Observations}~\label{sec: Observations}

In this section, we describe the important observations on how the sponsored ads may affect the quality of search results on Amazon. We perform our analyses across three important factors: (a)~How prevalent are sponsored results on Amazon SERPs for different queries? 
(b)~How relevant are the top sponsored results as per Amazon's own organic ranking? 
(c)~How comparable are the top sponsored results with the top organic results (based on price and quality)?.

\subsection{How prevalent are sponsored results on Amazon SERPs for different queries?}

For all queries across different snapshots, we observe some sponsored results interleaved among the organic results. 
Table~\ref{Tab: OrganicSponsored} shows the number of organic results and sponsored results collected across the different settings during our data collection. Across the board,  almost $1/3^\textrm{rd}$ of the observed Amazon SERPs constitute of sponsored results. Apart from France, the percentage of sponsored results increases in the default setting. 
However, the presentation of these sponsored results is not uniform. 
Usually, sponsored results are shown in the top, middle and at the end of a SERP depending on the number of results shown on the SERP.

\begin{figure}[t]
	\centering
	%\includesvg[width=\textwidth]{figures/Amazon-SERP-BabyWipes.svg}
            \includegraphics[width=0.6\columnwidth, height = 4 cm]{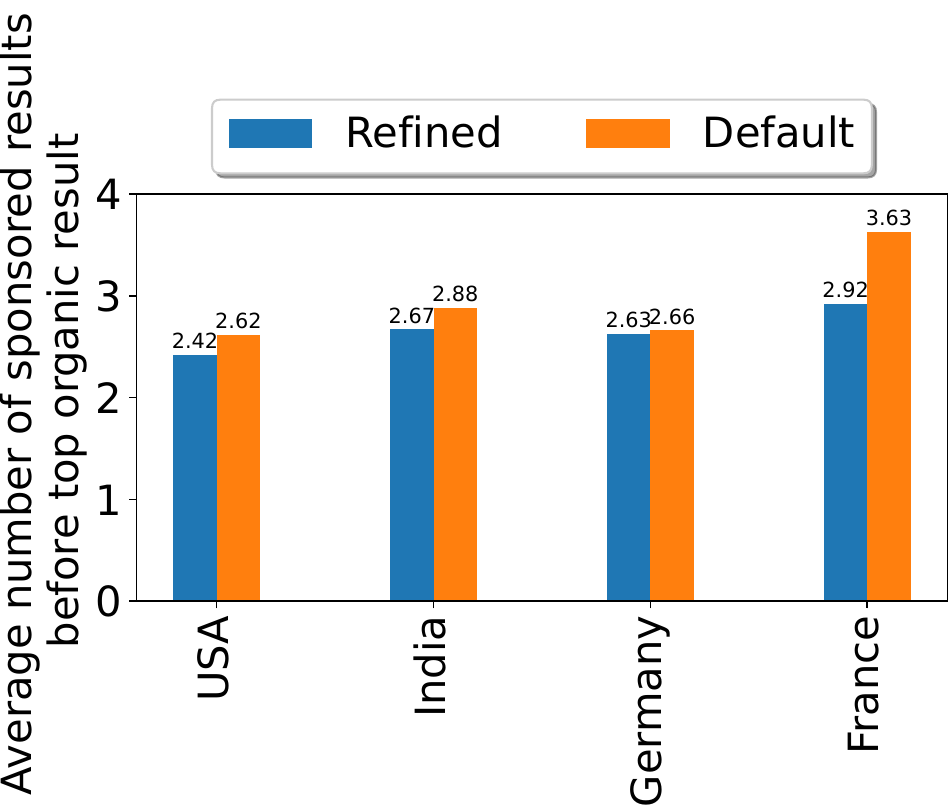}
         \caption{\textbf{Average number of sponsored results appearing before the top organic search result across the four marketplaces for all the queries in different settings. In general 2--3 sponsored results precede organic results; the number is slightly higher in default setting across the four marketplaces consistently.}}
    \label{Fig: SpBeforeOrg}
\end{figure}

For the sake of the analyses reported in this section, we will rely on \textit{sponsored results which especially appear before even the first organic result}. We focus on these sponsored results since they are almost certain to get the maximum attention of the consumers.
In the refined setting of our data collection for 50 queries across \textit{four} snapshots across the four marketplaces, we find that in 84.5\% queries, at least one sponsored result appear before the top organic result. 
The same percentage increases slightly to 87.37\% in the default setting (where a user queries the search box with `all categories' setting). This increase is due to a significant rise in the provision of ads in default setting for the US marketplace. 
The percentage for each of the marketplaces is shown in Table~\ref{Tab: prevalenceAds}. 
The average number of sponsored results appearing before top organic results is shown in Figure~\ref{Fig: SpBeforeOrg}. We observe that 2--3 sponsored results usually appear before the top organic result irrespective of the marketplace and search setting. The average number of sponsored result before top organic result, in fact, increases in the default setting. %across all marketplaces.

\vspace{1 mm}
\noindent
\textbf{Implications:} While it is very intuitive that Amazon SERPs show both sponsored and organic results to end-users, nearly 30\% of all results on Amazon SERP are observed to be sponsored. Interestingly, 2-3 sponsored results usually precede the top organic result on the first Amazon SERP for most of the searches we conducted. This observation is striking because in nearly 85\% (or above) cases, at least one sponsored result appears even before the top organic result on these SERPs. Note that, many end-users upon searching for a product tend to click or engage more with the first results that appear in the SERPs~\cite{dash2022alexa, Rubin2019Cracking}. Our observation shows that in 85\% cases, those first results are sponsored results.

\begin{table}
    \centering
    \small
    \begin{tabular}{|c|c|c|}
        \hline
         Country & \#Organic results & \#Sponsored results\\
         \hline \hline
         \multicolumn{3}{|c|}{Refined setting (50 queries $\times$ 4 snapshots $\times$ 4 marketplaces)}\\
         \hline
         USA & 143,000 & 40,177 (28.09\%) \\
         \hline
         India & 1,069,399 & 228,068 (21.32\%)\\
         \hline
         Germany & 97,930 & 29,532 (30.16\%)\\
         \hline
         France & 117,899 & 33,260 (28.21\%)\\
        \hline \hline
        \multicolumn{3}{|c|}{Default setting (1000 queries $\times$ 1 snapshot $\times$ 4 marketplaces)}\\
        \hline
        USA & 258,446 & 84,951 (32.87\%) \\
         \hline
         India & 230,432 & 64,881 (28.15\%)\\
         \hline
         Germany & 272,978 & 92,646 (33.94\%)\\
         \hline
         France & 240,510 & 64,545 (26.84\%)\\
         \hline
    \end{tabular}
    \caption{\textbf{Number of organic and sponsored results collected in each of the data collection setting. Almost $1/3^{rd}$ of Amazon SERP are sponsored results.}}
    \label{Tab: OrganicSponsored}
\end{table}

\begin{table}
    \centering
    \begin{tabular}{|c|c|c|}
        \hline
        Marketplace & Refined & Default\\
         \hline \hline
         USA & 77.0\% & 91.5\%\\
         \hline
         India & 83.5\% & 77.7\%\\
         \hline
         Germany & 89\% & 91.9\%\\
         \hline
         France & 88.5\% & 88.4\%\\
         \hline
         Overall & 84.5\% & 87.37\% \\
         \hline
    \end{tabular}
    \caption{\textbf{Percentage of queries where we found a sponsored result (advertisement) appearing even before the top organic result across the different Amazon marketplaces in the two settings of data collection.}}
    \label{Tab: prevalenceAds}
\end{table}

\subsection{How relevant are the top sponsored results to the query, as per Amazon's own ranking?}

The previous observation demonstrates that a significant fraction of Amazon SERPs show sponsored results before the top organic results. 
The key follow up question is -- how relevant are the sponsored results to the query? 
If the sponsored results (that appear before the organic results) are actually very relevant to the queries, then they may not necessarily degrade the search result quality. In fact in such cases, putting such sponsored results may lead to consumer satisfaction, while fostering competition by giving exposure to other competing sellers.

\noindent
\textbf{Relevance of sponsored results:} There are multiple ways to operationalise the notion of relevance for different results in the context of a query. 
One can come up with some simple model that utilises the different publicly available features of the products to rank them in decreasing order of relevance.  
To this end, we consider the Amazon search ranking algorithm itself,  which leverages multiple nuanced features (some public and some private) to come up with the final ranking of products based on relevance. Hence, we choose to collect Amazon search results till the end (or till allowed), to see where a sponsored result appears organically. In other words, the rank at which a sponsored product appears for the first time as an organic result is considered as a proxy for its relevance (as evaluated by Amazon's ranking algorithm). 

\begin{figure*}[t]
	\centering
	\begin{subfigure}{0.24\textwidth}
		%\centering
		\includegraphics[width= \textwidth, height=3.75cm]{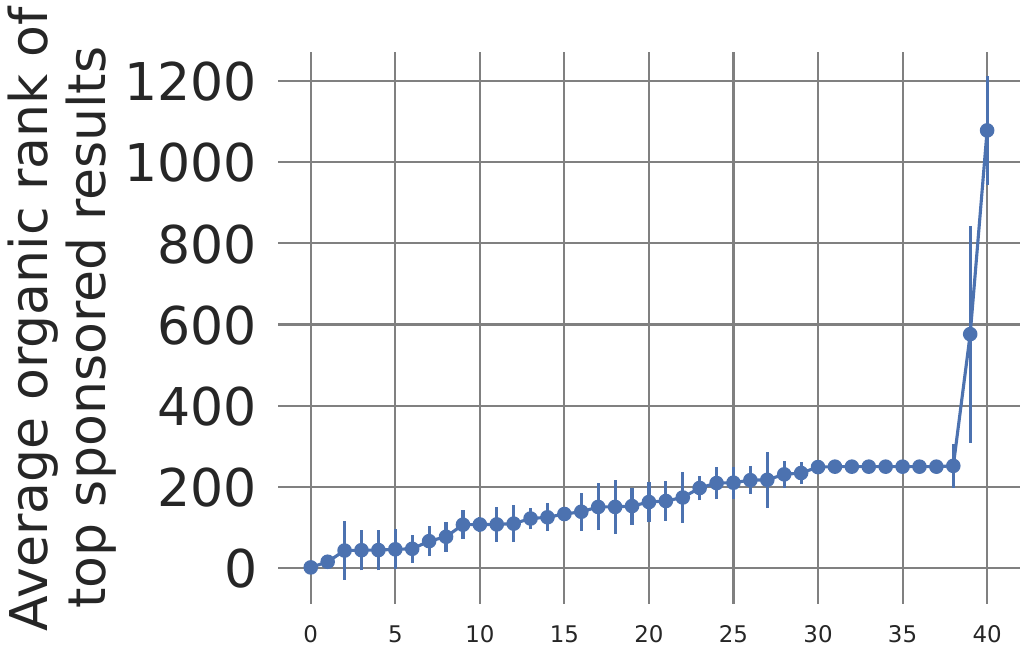}
		%\vspace*{-4mm}
		\caption{USA (Refined)}
		%\label{Fig: buybox2}
	\end{subfigure}%
        \begin{subfigure}{0.24\textwidth}
		%\centering
		\includegraphics[width= \textwidth, height=3.75cm]{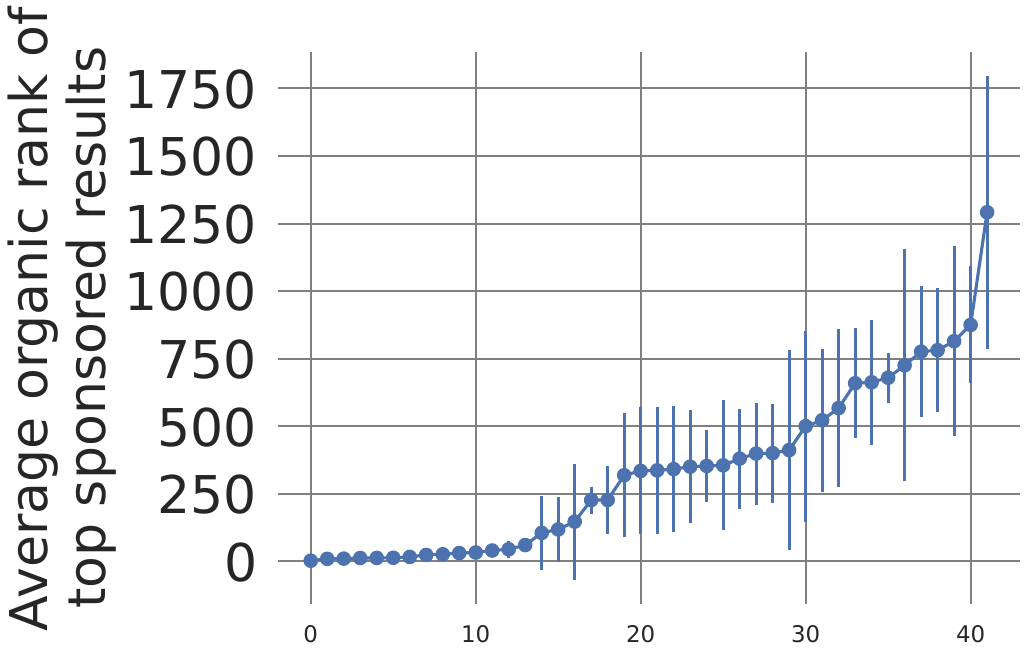}
		%\vspace*{-4mm}
		\caption{India (Refined)}
		%\label{Fig: buybox2}
	\end{subfigure}%
        \begin{subfigure}{0.24\textwidth}
		%\centering
		\includegraphics[width= \textwidth, height=3.75cm]{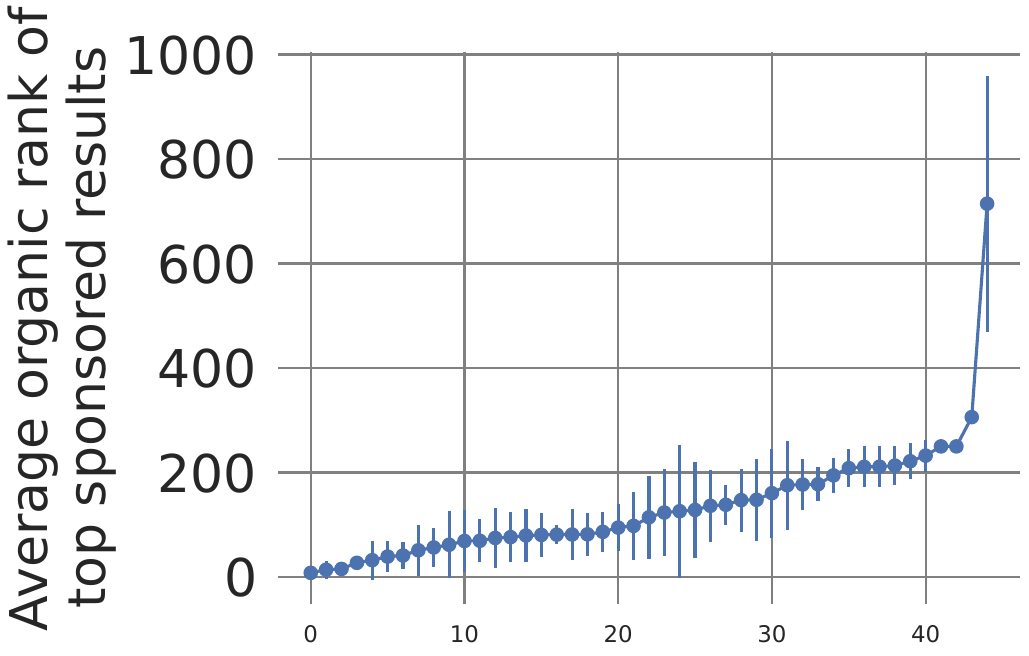}
		%\vspace*{-4mm}
		\caption{Germany (Refined)}
		%\label{Fig: buybox2}
	\end{subfigure}%
        \begin{subfigure}{0.24\textwidth}
		%\centering
		\includegraphics[width= \textwidth, height=3.75cm]{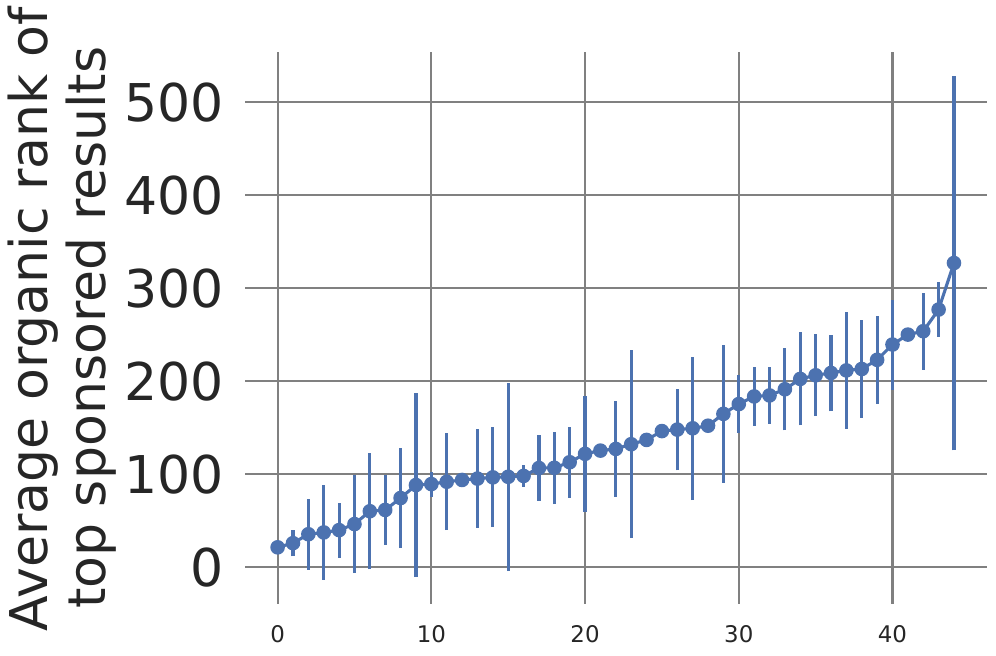}
		%\vspace*{-4mm}
		\caption{France (Refined)}
		%\label{Fig: buybox2}
	\end{subfigure}%

        \begin{subfigure}{0.24\textwidth}
		%\centering
		\includegraphics[width= \textwidth, height=3.75cm]{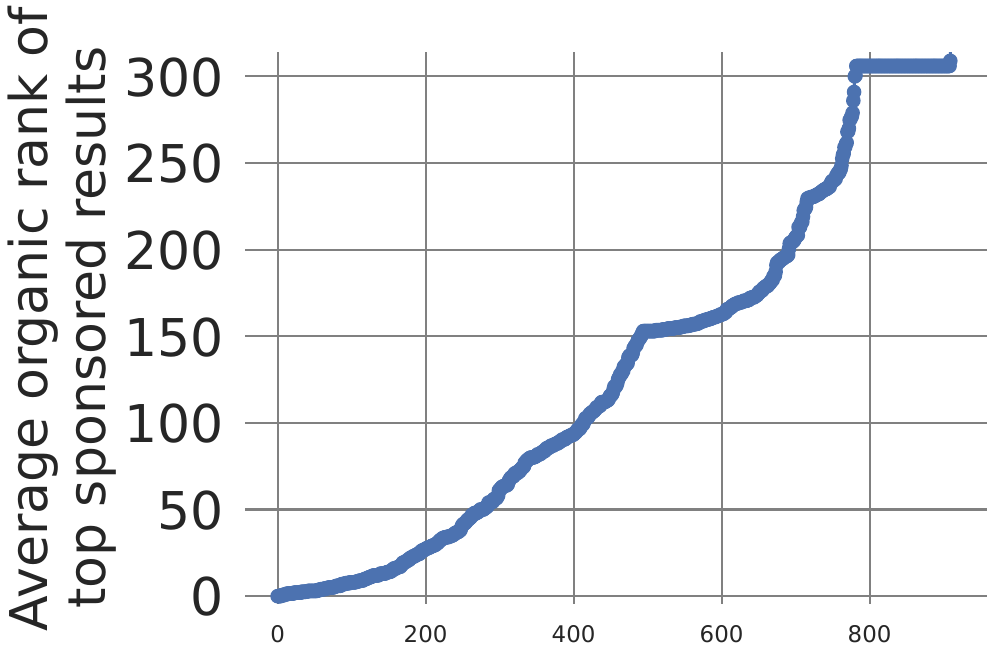}
		%\vspace*{-4mm}
		\caption{USA (Default)}
		%\label{Fig: buybox2}
	\end{subfigure}%
        \begin{subfigure}{0.24\textwidth}
		%\centering
		\includegraphics[width= \textwidth, height=3.75cm]{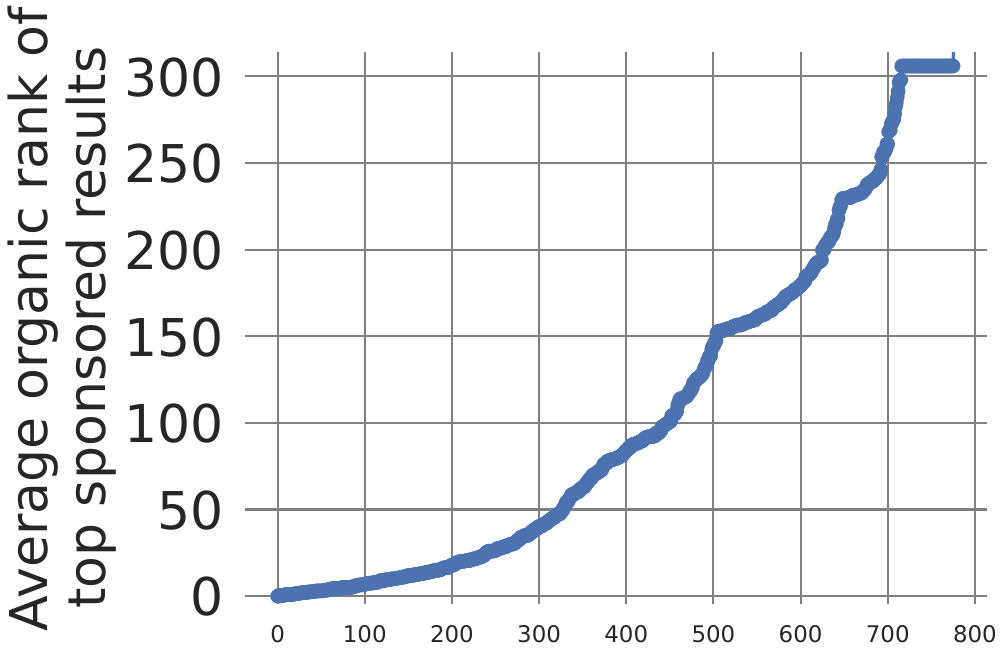}
		%\vspace*{-4mm}
		\caption{India (Default)}
		%\label{Fig: buybox2}
	\end{subfigure}%
        \begin{subfigure}{0.24\textwidth}
		%\centering
		\includegraphics[width= \textwidth, height=3.75cm]{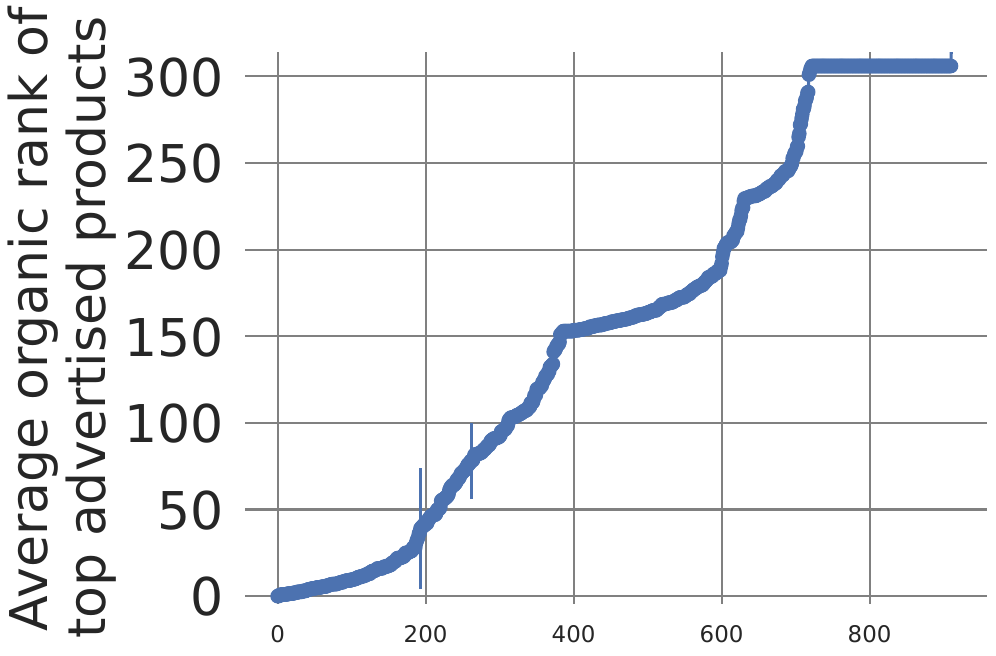}
		%\vspace*{-4mm}
		\caption{Germany (Default)}
		%\label{Fig: buybox2}
	\end{subfigure}%
        \begin{subfigure}{0.24\textwidth}
		%\centering
		\includegraphics[width= \textwidth, height=3.75cm]{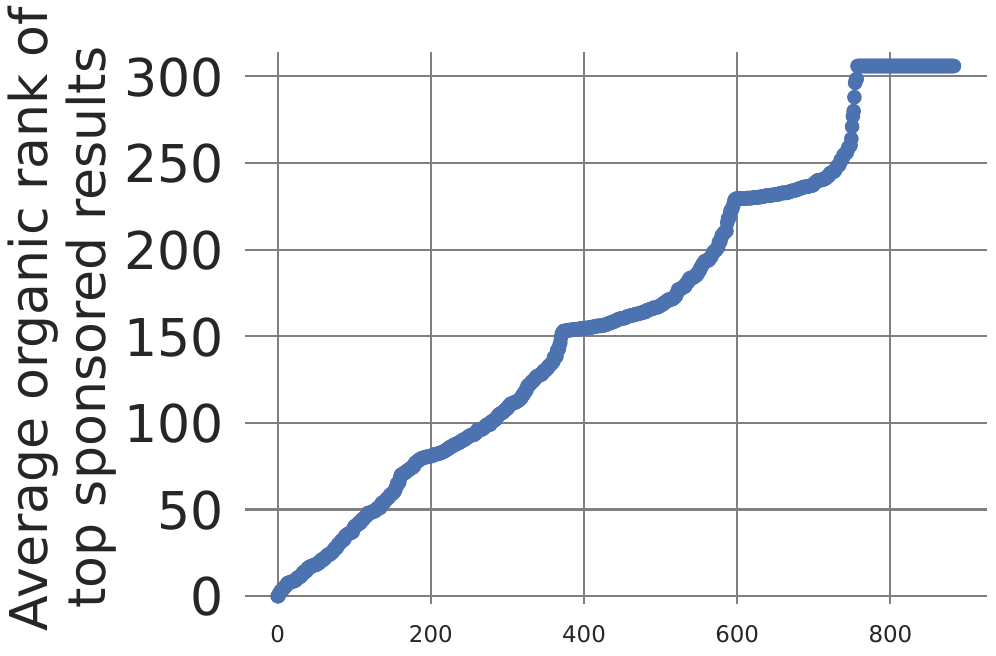}
		%\vspace*{-4mm}
		\caption{France (Default)}
		%\label{Fig: buybox2}
	\end{subfigure}%
	%\vspace{-3mm}
	\caption{ {\bf Mean (and standard deviation error bar) organic rank, as evaluated by Amazon's own search ranking algorithm, of top sponsored results appearing before the top organic results on Amazon SERPs for different queries (along X-axis) across the four marketplaces. For a large majority of cases the top sponsored results appear below organic rank 100 across the four marketplaces for different queries consistently.}
	}
	\label{Fig: OrgRank}
    \vspace{-4 mm}
\end{figure*}

\begin{table}
    \centering
    \begin{tabular}{|c|c|c|c|}
        \hline
        Countries & $1^\textrm{st}$ page & $2^\textrm{nd}$ page & $3^\textrm{rd}$ page\\
        \hline \hline
        \multicolumn{4}{|c|}{Refined setting (50 queries $\times$ 4 snapshots)}\\
        \hline
        USA & 73.44\% & 67.90\% & 64.55\%\\
        \hline
        India & 57.55\% & 49.63\% & 43.68\%\\
        \hline
        Germany & 65.09\% & 53.15\% & 48.38\%\\
        \hline
        France & 63.43\% & 57.87\% & 52.55\%\\
        \hline \hline
        \multicolumn{4}{|c|}{Default setting (1000 queries)}\\
        \hline
        USA & 61.45\% & 51.35\% & 46.29\%\\
        \hline
        India & 54.59\% & 43.18\% & 38.21\%\\
        \hline
        Germany & 67.08\% & 58.49\% & 54.18\%\\
        \hline
        France & 64.98\% & 55.26\% & 48.97\%\\
        \hline
    \end{tabular}
    \caption{\textbf{Percentage of top sponsored results on Amazon SERPs appearing in organic positions that are beyond the first, second and third SERPs respectively. Apart from India, in all the other marketplaces more than 50\% of sponsored results appear beyond the second SERP for the first time as an organic result.}}
    \label{Tab: BeyondPage}
\end{table}

\noindent
\textbf{Where do the top sponsored results appear organically?} 
Figure~\ref{Fig: OrgRank} show the average organic rank of top sponsored results for different queries in our collected dataset. For example, for a query \textit{q}, let us assume 3 sponsored results appear before the top organic result, and the organic ranks of those products are at positions $x, y, z$. Then the corresponding Y-axis value shows the mean of the three organic positions i.e., $(x+y+z)/3$. 
The higher the value, lesser the relevance -- i.e., 0 is the top ranked most relevant product as per Amazon's organic ranking.
If a sponsored product is \textit{not} found at all in the organic ranked list, we consider its organic position to be the total length of the organic result list.
Figure~\ref{Fig: OrgRank} (a--d) show the average organic rank of top sponsored results along with a standard deviation error bar (variation across 4 temporal snapshots) on Y-axis for the different queries (X-axis).  
The average organic rank of most SERPs are found to be beyond 100 i.e., the overall relevance of the top sponsored results is not comparable with the top organic results in the observed search results. 
Similarly, Figure~\ref{Fig: OrgRank} (e--h) show the same observation for the default setting i.e., where a total of 306 organic results are shown. Much like the observation in the refined setting, in the default setting as well the top sponsored results often come from very bad ranks as per Amazon's organic search ranking.

Note that nearly 70\% of consumers never go beyond the first Amazon search engine result page~\cite{FTC2023Federal}. If we consider a geometric decay of exposure, then nearly 97.3\% consumers would never go beyond the third page of Amazon's search results. If we analyse the organic ranks of the advertised products closely, we see that a large majority of them do not get positioned in the top-3 Amazon SERPs as organic results i.e., they are significantly less relevant as per Amazon's own search ranking. Table~\ref{Tab: BeyondPage} shows the fraction of times the top sponsored results appear beyond the top-3 Amazon SERPs in our observed SERPs. 
For example, in the United States 64.55\% of the top sponsored products do not get organic positions among the first 3 Amazon SERPs. 
In general, nearly 50\% of sponsored results do not merit a position in the top-3 SERPs based on Amazon's ranking system.
Similar observations are made in the default setting. However, we notice a drop in percentages in the default setting as compared to the refined setting (apart from Germany).

\subsection{Are the top sponsored results under-explored?}
\begin{table}
    \centering
    \begin{tabular}{|c|c|c|}
        \hline
         Country & $\ge 1000 $ \#Ratings & $\ge 5000 $ \#Ratings\\
         \hline \hline
         \multicolumn{3}{|c|}{Refined setting (50 queries $\times$ 4 snapshots)}\\
         \hline
         USA & 57.49 & 42.51 \\
         \hline
         India & 60.09 & 31.61\\
         \hline
         Germany & 52.99 & 26.28\\
         \hline
         France & 45.95 & 26.28\\
        \hline \hline
        \multicolumn{3}{|c|}{Default setting (1000 queries)}\\
        \hline
        USA & 47.58 & 30.80 \\
         \hline
         India & 42.58 & 22.47\\
         \hline
         Germany & 46.19 & 23.26\\
         \hline
         France & 44.82 & 22.74\\
         \hline
    \end{tabular}
    \caption{\textbf{Percentage of times top sponsored results have \#Ratings more than 1000 and 5000 respectively.}}
    \label{Tab: PopularityThreshold}
    %\vspace{-2 mm}
\end{table}

Increasingly Amazon has promoted its sponsored product advertisements as a means to provide exposure to products (and their sellers) who may not have gotten exposure organically~\cite{Amazon2024Sponsored}.
Note that Amazon SERPs show the number of ratings that each product has received over their lifetime; which can be considered as a proxy for how many consumers may have purchased the same product.
Thus, we take a look at the number of ratings that the different sponsored products might have had.

The observations are listed in Table~\ref{Tab: PopularityThreshold}. We find that nearly 50\% of all top sponsored products have more than 1000 number of ratings i.e., a minimum of 1000 people have rated and reviewed those products. 
More than 50\% of those products in fact have the number of ratings beyond 5000.
Prior consumer surveys show that many consumers are more likely to purchase products that have more than 1000  ratings~\cite{Fischer2024Reviews}. 
Furthermore, the conversion rate drastically increases when the number of ratings go beyond 5000. Note that the median number of ratings received by \textit{all} products that we observed in our data collection across India, USA, Germany and France are 8, 57, 51, 29 respectively.
Hence, we can infer that the products appearing as top sponsored results  
are typically \textit{more explored} as opposed to the median products in our dataset. 
At the same time, they are more likely to be clicked by consumers because of higher \#Ratings because of their positioning before the top organic results.

\subsection{How comparable are the top sponsored and organic results on Amazon SERPs?}

Although many of the sponsored results we observed come from very low ranks, i.e., they are less relevant as per the Amazon's product ranking, we now check if they are comparable with the top organic products in terms of different features which consumers perceive to be important. For example, they may still be of comparable or less price; higher quality. In such cases, the eventual harm to consumers may not be as significant either.

%\noindent
%\textbf{Experimental setup: }
To fairly compare the sponsored and organic results, we turn our focus to the sponsored results appearing before the top organic results. For a query $q$, let us assume there are $k$ sponsored results appearing before the first organic result. In such case, we compare the top-$k$ sponsored results with the top-$k$ organic results on the basis of some property $p$ (e.g., price or quality). In such cases, we compare the average property of the top sponsored and organic results e.g., the average price of top-$k$ sponsored results and that of top-$k$ organic results for a particular query are compared. 

\begin{figure*}[t]
	\centering
	\begin{subfigure}{0.24\textwidth}
		%\centering
		\includegraphics[width= \textwidth, height=3.75cm]{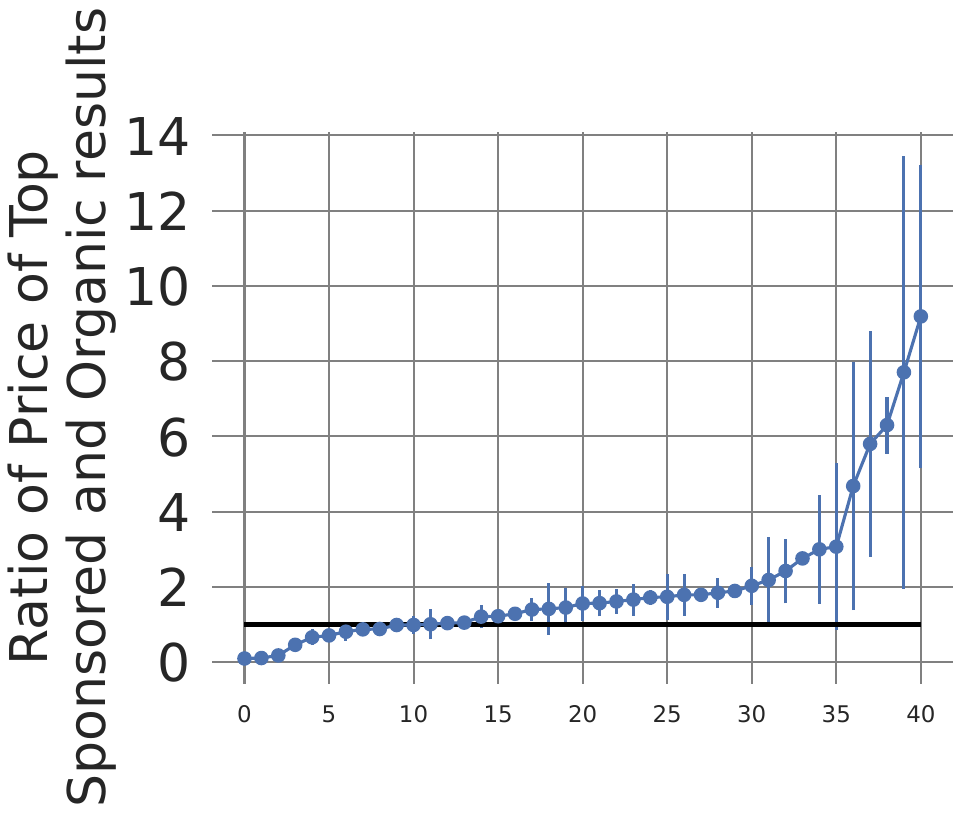}
		%\vspace*{-4mm}
		\caption{USA (Refined)}
		%\label{Fig: buybox2}
	\end{subfigure}%
        \begin{subfigure}{0.24\textwidth}
		%\centering
		\includegraphics[width= \textwidth, height=3.75cm]{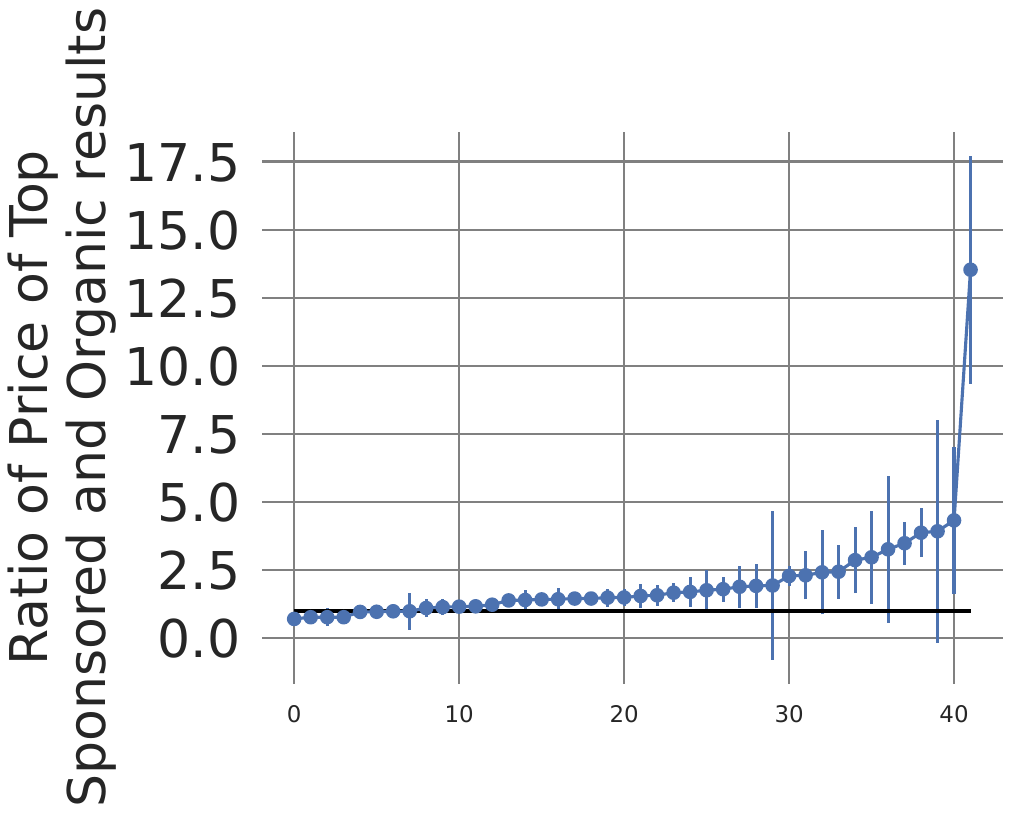}
		%\vspace*{-4mm}
		\caption{India (Refined)}
		%\label{Fig: buybox2}
	\end{subfigure}%
        \begin{subfigure}{0.24\textwidth}
		%\centering
		\includegraphics[width= \textwidth, height=3.75cm]{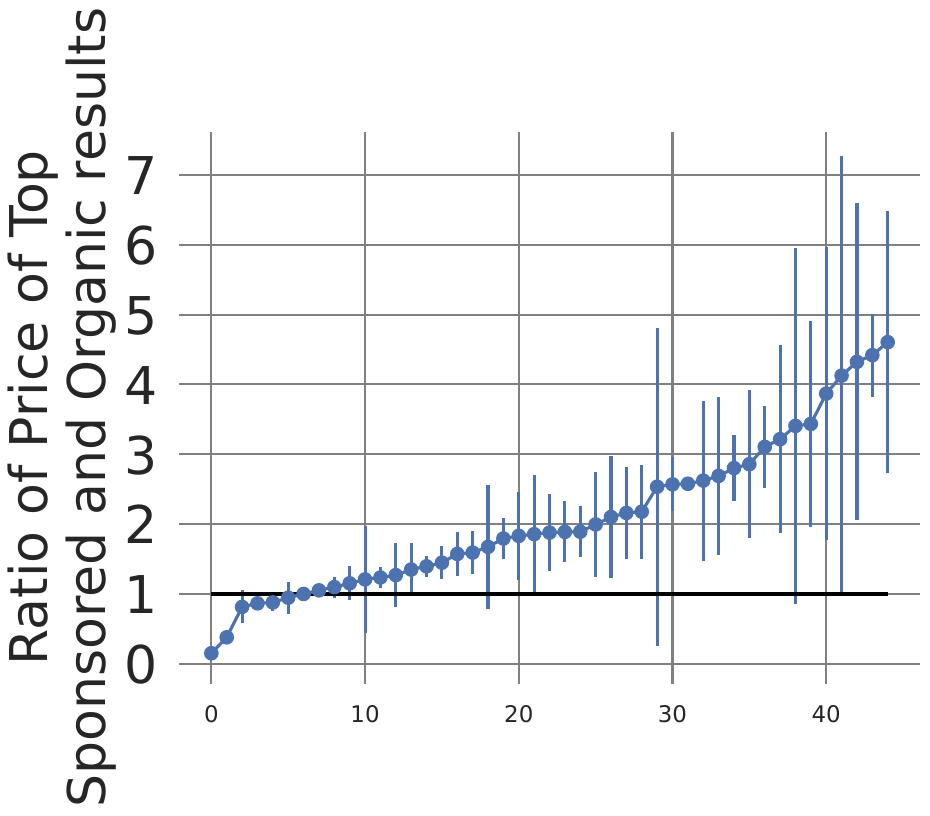}
		%\vspace*{-4mm}
		\caption{Germany (Refined)}
		%\label{Fig: buybox2}
	\end{subfigure}%
        \begin{subfigure}{0.24\textwidth}
		%\centering
		\includegraphics[width= \textwidth, height=3.75cm]{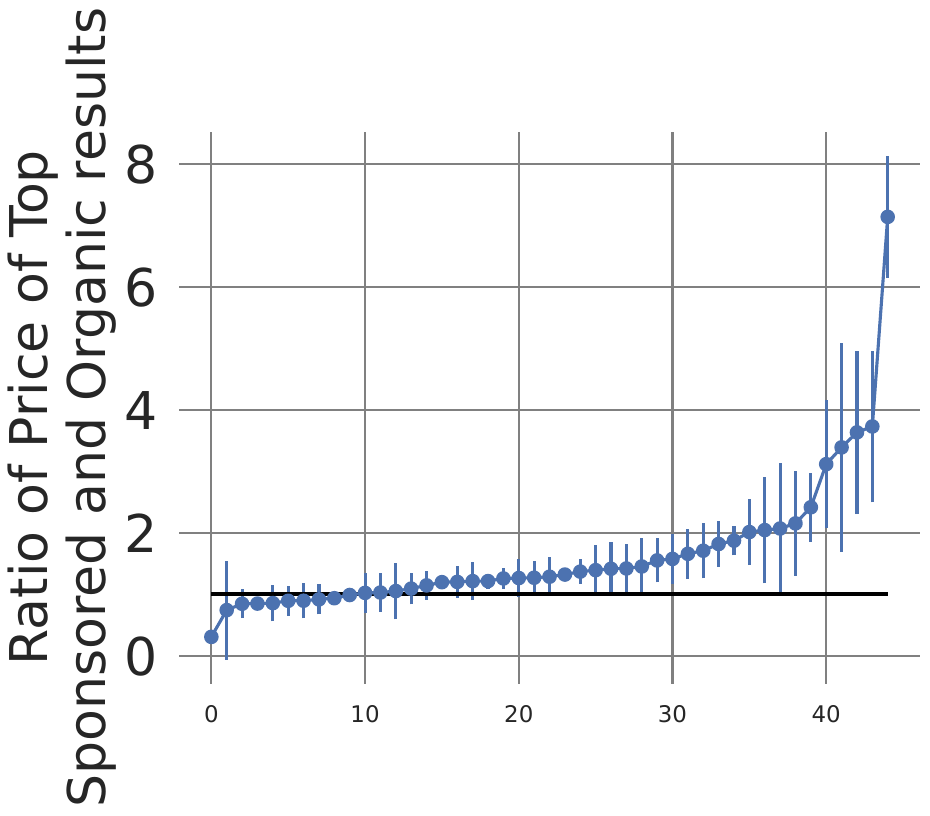}
		%\vspace*{-4mm}
		\caption{France (Refined)}
		%\label{Fig: buybox2}
	\end{subfigure}%

        \begin{subfigure}{0.24\textwidth}
		%\centering
		\includegraphics[width= \textwidth, height=3.75cm]{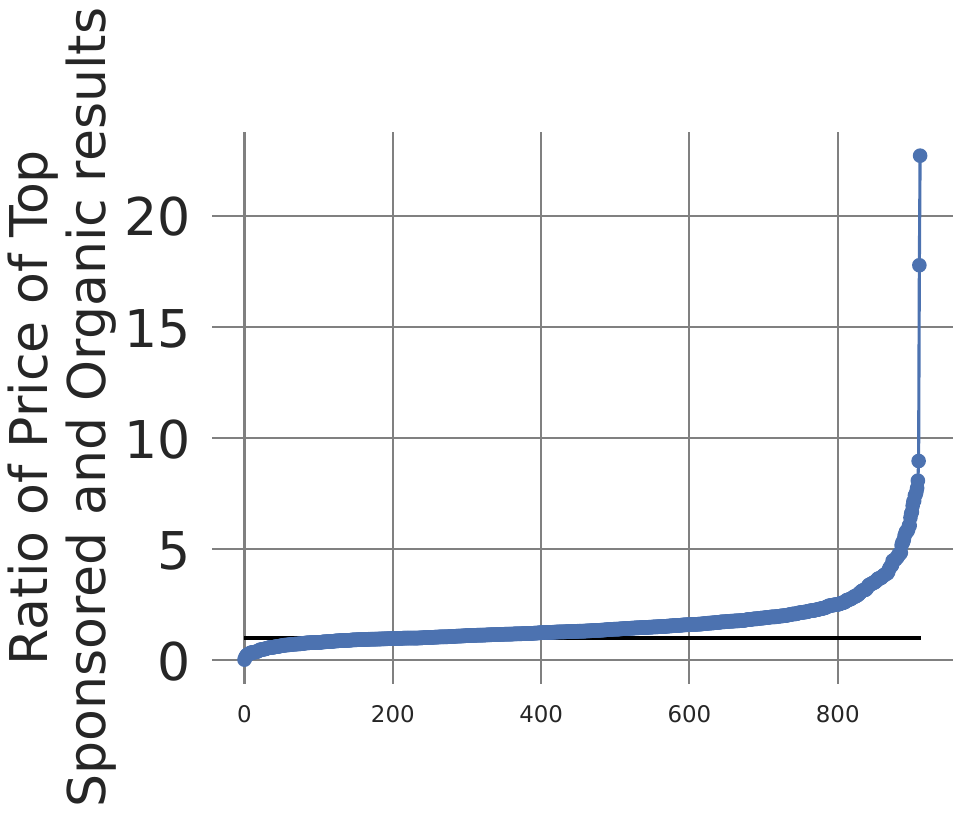}
		%\vspace*{-4mm}
		\caption{USA (Default)}
		%\label{Fig: buybox2}
	\end{subfigure}%
        \begin{subfigure}{0.24\textwidth}
		%\centering
		\includegraphics[width= \textwidth, height=3.75cm]{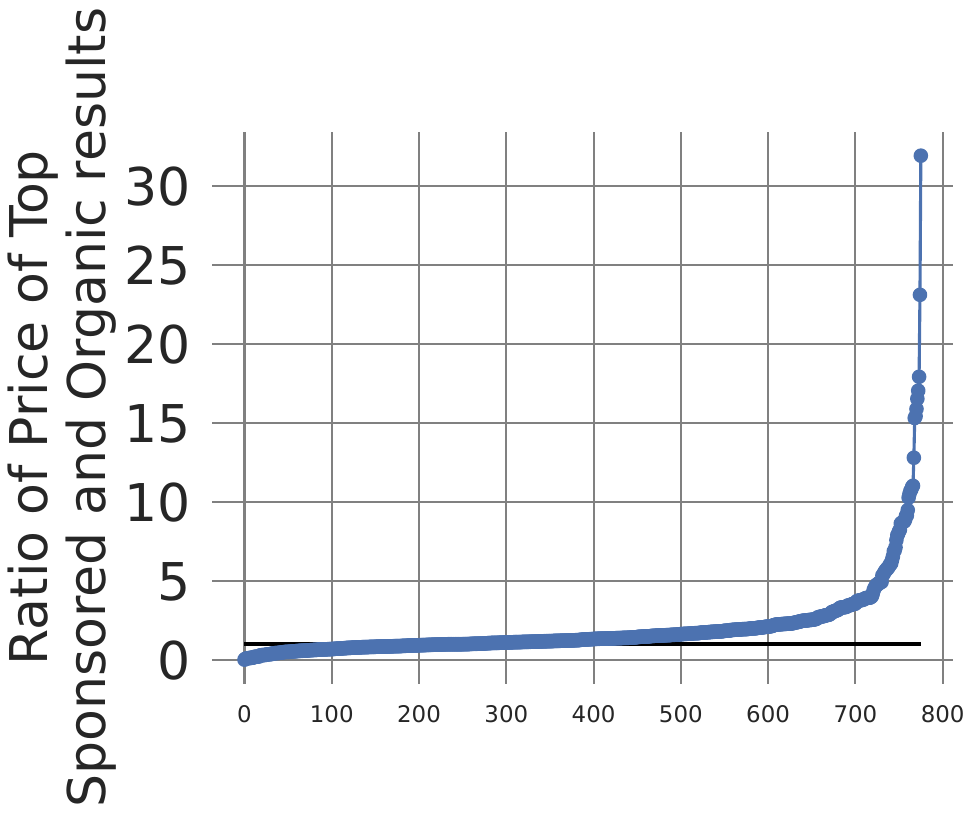}
		%\vspace*{-4mm}
		\caption{India (Default)}
		%\label{Fig: buybox2}
	\end{subfigure}%
        \begin{subfigure}{0.24\textwidth}
		%\centering
		\includegraphics[width= \textwidth, height=3.75cm]{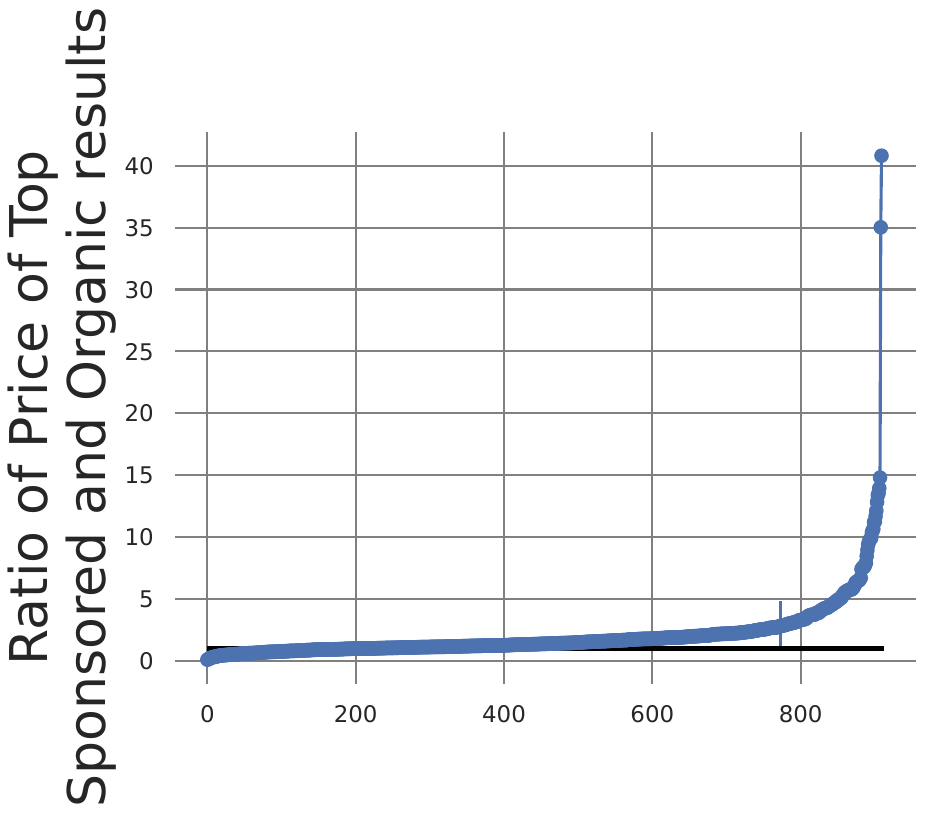}
		%\vspace*{-4mm}
		\caption{Germany (Default)}
		%\label{Fig: buybox2}
	\end{subfigure}%
        \begin{subfigure}{0.24\textwidth}
		%\centering
		\includegraphics[width= \textwidth, height=3.75cm]{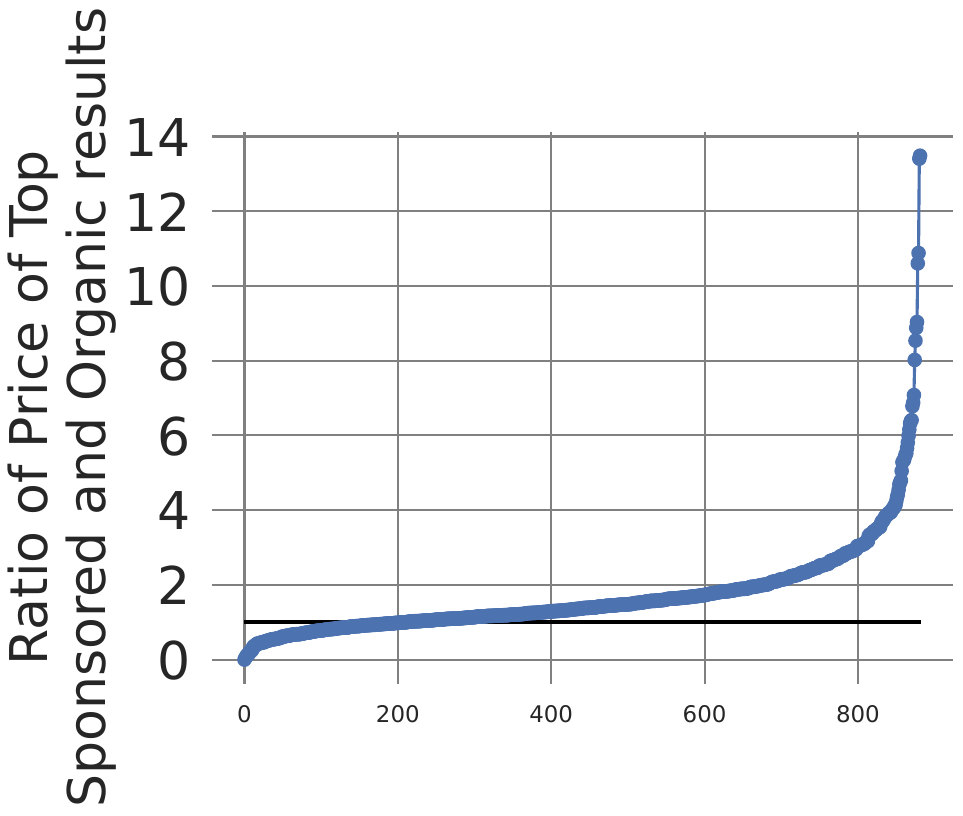}
		%\vspace*{-4mm}
		\caption{France (Default)}
		%\label{Fig: buybox2}
	\end{subfigure}%
	%\vspace{-3mm}
	\caption{ {\bf Mean (and standard deviation error bar) ratio of average price of top sponsored and organic results for different queries (along X-axis) on Amazon SERP. For a large majority of cases the top sponsored results are costlier than top organic results.}
	}
    \label{Fig: PriceCompare}
    \vspace{-4 mm}
\end{figure*}

\noindent
\textbf{Top sponsored results are costlier than top organic results: }
Price is one of the key features which helps consumer in transactional decision making. Hence, first we check how comparable are the prices of top sponsored and organic results on Amazon SERPs we collected. 
Figure~\ref{Fig: PriceCompare} shows the mean and standard deviation error bar of ratio of price of top sponsored results to top organic results for all the queries where there are at least one sponsored result shown before the organic result on the Y-axis. 
We have highlighted the horizontal line at $Y = 1$ as a baseline for the average prices to be similar.
We observe that the top sponsored products are costlier in 70.78\%, 77.24\%, 80.89\% and 74.01\% cases, in USA, India, Germany and France respectively than the organic products in our collected SERPs.

Table~\ref{Tab: PricePercentage} shows the difference between the average prices. For example, in more than 50\% cases, the average price of the top sponsored results is more than 1.5 times that of average price of as many top organic results. Similar trends are observed in the default setting, a setting which is used by millions of consumers. In fact, in almost $1/3^\textrm{rd}$ of the cases in the collected SERPs, the average price of top sponsored results is more than double that of the top organic results.

\begin{table}
    \centering
    \begin{tabular}{|c|c|c|c|}
        \hline
        Countries & $\ge 1.2\times$ & $\ge 1.5\times$ & $\ge 2\times$\\
        \hline \hline
        \multicolumn{4}{|c|}{Refined setting (50 queries $\times$ 4 snapshots)}\\
        \hline
        USA & 80.73\% & 63.30\% & 39.45\%\\
        \hline
        India & 79.84\% & 54.26\% & 37.98\%\\
        \hline
        Germany & 86.81\% & 70.14\% & 49.31\%\\
        \hline
        France & 82.44\% & 48.09\% & 29.01\%\\
        \hline \hline
        \multicolumn{4}{|c|}{Default setting (1000 queries)}\\
        \hline
        USA & 78.53\% & 53.23\% & 27.50\%\\
        \hline
        India & 80.45\% & 59.39\% & 37.19\%\\
        \hline
        Germany & 80.11\% & 58.65\% & 36.19\%\\
        \hline
        France & 80.18\% & 55.89\% & 30.70\%\\
        \hline
    \end{tabular}
    \caption{\textbf{Percentage of times when top sponsored results on Amazon are 1.2, 1.5 or 2$\times$ costlier than the top organic results. More than 50\% cases, the average price of the top sponsored results is more than 1.5$\times$ of the average price of the top organic results.}}
    \label{Tab: PricePercentage}
\end{table}

\noindent 
\textbf{Quality of organic results are superior: }Much like price, the quality of the product is of utmost importance in consumers' decision making. For our experiment, we take a product's average user rating (out of 5 stars) as an indicator of its quality. We observe that the average quality of top sponsored products are less than that of the organic products in 43.5\%, 47.3\%, 57.3\% and 59.89\% respectively in USA, India, Germany and France (see Table~\ref{Tab: QualityCompare}). The observations are qualitatively similar for the default setting as well across the different marketplaces.

\begin{table}
    \centering
    \begin{tabular}{|c|c|c|}
        \hline
         Country & Better quality & Lower quality\\
         \hline \hline
         \multicolumn{3}{|c|}{Refined setting (50 queries $\times$ 4 snapshots)}\\
         \hline
         USA & 56.50 & 43.50 \\
         \hline
         India & 52.70 & 47.30\\
         \hline
         Germany & 42.70 & 57.30\\
         \hline
         France & 40.11 & 59.89\\
        \hline \hline
        \multicolumn{3}{|c|}{Default setting (1000 queries)}\\
        \hline
        USA & 50.71 & 49.29 \\
         \hline
         India & 52.50 & 47.50\\
         \hline
         Germany & 42.76 & 57.24\\
         \hline
         France & 41.41 & 58.59\\
         \hline
    \end{tabular}
    \caption{\textbf{Percentage of times top sponsored results are of better or comparable quality than the top organic results in its vicinity. Sponsored results are not of any better quality than the organic results.}}
    \label{Tab: QualityCompare}
\end{table}

\subsection{Implications of our observations}

\vspace{1 mm}
\noindent
\textbf{Implications for the consumers: } If a product appearing at organic position 100 is shown to a consumer as the first result (albeit sponsored), consumers are more likely to click on such results due to their cognitive position biases~\cite{baeza2018bias}. In fact, Amazon itself reports that nearly 35\% consumers click on the first result appearing on Amazon SERPs~\cite{Rubin2021Cracking}; which increasingly is a sponsored result or an advertisement. 
Further, one should also note that \citet{Graham2019Amazon} showed almost half of the participants on Amazon cannot even distinguish between sponsored and organic results. Thus, positioning such irrelevant product on top of observed rankings may potentially nudge consumers to significantly less relevant options for their desired query leading to consumer dissatisfaction. 

While positioning irrelevant products decreases search quality, it also is observed to nudge consumers toward potentially costlier products. While Amazon is known for its consumer-centric policies, this comes as a contradiction. 
Even worse, these products are not even of significantly better quality i.e., even the quality cannot always justify the extra cost that consumers have to pay.

\vspace{1 mm}
\noindent
\textbf{Implications for the sellers and competition: }
From the perspective of sellers, a number of sellers who have struggled hard to achieve the top organic positions based on prior engagement / purchase etc., are downgraded (below sponsored results) on the observed search ranking. This, in turn, may deprive them of opportunities to sales and revenue. 

Furthermore, such practices can have bigger implications on the competition in the marketplaces. By virtue of promoting sponsored results at the cost of organic results, essentially sellers are encouraged to invest more on advertising than to improve or maintain product quality and / or competitive prices. In other words, increasingly advertisement has become the primary gateway for business users to reach and attract the attention of the large consumer base of Amazon.
This not only increases the costs that sellers must bear to reach potential consumers but also results in those extra costs being passed down to the consumers resulting in higher price.
Further, this may even hand in an advantage to potentially large sellers on marketplace who can afford to spend big on advertising. 
\section{Legal implications of our observations}\label{Sec: Legal}
Policymakers across the globe have been proactive on preserving fairness in commercial practices especially from the vantage points of the businesses and consumers. In this section, we briefly discuss the observations in context of some of the consumer protection and fair commercial practice guidelines in the relevant parts of the world. 

\subsection{The United States of America}
Section 5 of the Federal Trade Commission (FTC) Act declared `unfair methods of competition in or affecting commerce, and unfair or deceptive acts of practices in or affecting commerce' to be unlawful~\cite{FTC2006Federal}. 
Usually, the severity of the practice matters a lot in enforcement of this regulation. Because of the ex-post nature of the above guidelines, different practices of large digital marketplace players have avoided legal actions on the basis of their potential benefits toward consumers i.e., consumer welfare~\cite{Sallet2018Competitive}.
Note that advertising is not an unfair trade practice in and of itself. However, the competitive nature and the potential of distorting organic consumer behavior at scale is the biggest concern in the context of interleaving advertisements within search results in digital marketplaces.
Marking a shift in the policymakers' approach, recently, Amazon has come under intense scrutiny in the USA for its business practices ranging from manufacturing private labels to preferentially treating its preferred sellers (or itself)~\cite{Amazon2019Online,Amazon2020Questions}. Recently, US FTC has also filed a law-suit against Amazon for its unfair trade practices~\cite{FTC2023Federal} which mentions advertisement on Amazon search among other things.

Specifically, the FTC lawsuit alleges Amazon's advertising policies are in direct violation of Consumer Fraud Act of New Jersey state~\cite{NJ2022Fraud}. This law prohibits adopting any unconscionable commercial practice, deception in connection with sale or advertisement. Amazon's degradation of quality of shopper-facing search results and nudging customers to more expensive items, as per the FTC lawsuit, do not comply with the said regulation.

\noindent
\textbf{Potential non-compliance on Amazon: }The recent US FTC law suit sheds some light on internal studies of Amazon where they find the increased advertising revenue outweighs the sales it loses from worsening the relevance and quality of search results~\cite{FTC2023Federal}. In other words, such practices are profitable at the cost of consumer injury. Practices of interleaving and even highlighting sponsored results at the cost of organic results may result in substantial consumer harm by degrading the overall consumer experience by nudging consumers to irrelevant, costlier and even lower quality products, as we showed in the previous section. Such harms do not even outweigh the potential exposure it may provide to less popular products and their sellers. Furthermore, since the SERP interface and the ranking algorithm mediates the interaction between consumers and shown products at a massive scale, such degradation is arguably unavoidable by consumers unless they are proactive and attentive.

\subsection{India}
Section 2(47) in Indian Consumer Protection Act also addresses adoption of any unfair method or deceptive practice to be unlawful for the purpose of promoting the sale~\cite{DCA2019Consumer}. In 2023, Central Consumer Protection Authority further provided a set of guidelines, especially in the context of digital marketplaces, to prevent and regulate \textit{dark patterns}~\cite{DCA2019Consumer}. Dark pattern is defined in the said notification as, `any practices or deceptive design pattern using user interface or user experience interactions on any platform that is designed to mislead or trick users to do something they originally did not intend or want to do'. The set of potential dark pattern also included \textit{disguised advertisement} and \textit{interface interference}. Under the first point, the guideline prohibits the practice of posing, masking advertisements as other types of contents blending with the rest of an interface to trick consumers into clicking them. The second point prohibits manipulating consumers in user interfaces by highlighting specific information to affect their desired action. 

\noindent \textbf{Potential non-compliance on Amazon: } We believe the interleaving of sponsored results among organic search result, and especially putting them above even the top organic results may not be desirable under such guidelines. In particular, such practices of blending ads with other types of (organic) results potentially constitutes a dark pattern (disguised advertisements). These practices may materially distort a consumers' decision making as many of them tend to click top results~\cite{Rubin2019Cracking}.

\subsection{The European Union}
Article 5 of the \textit{Unfair Commercial Practices Directive} of the European Union defines an unfair commercial practice as one which materially distorts or is likely to materially distort the economic behavior with regard to the product of average consumer whom it reaches. It further prohibits any such commercial practices~\cite{EU2021Unfair}. Article 8 of the same directive, further prohibits aggressive commercial practices which can likely manipulate the transaction decisions of average consumers. Article 9 also prohibits exploitation of any specific misfortune or circumstances to impair or influence consumers' decision with regard to the product. 

Going one step further, 
the EU has recently come up with a set of \textit{ex-ante} regulations for digital platforms in the form of Digital Services Act (DSA)~\cite{EU2022DSA} and Digital Markets Act (DMA)~\cite{EC2022DMA}. 
The primary difference in these \textit{ex-ante} regulations from the prior regulations 
is here the rules and obligations are established before the unfairness concerns arise. Therefore, all stakeholders have to obey them once they come into force.  
Specifically, the Article 25 of DSA prohibits online platforms from using deceptive design choices in their interfaces which may distort the end users' ability to make a free and informed decision.

\noindent \textbf{Potential non-compliance on Amazon:}
As we have shown earlier, the advertisements in Amazon SERPs are being put even before the first organic search result, which may materially distort/manipulate the consumers' economic behavior and transaction decisions especially toward the top sponsored results (e.g., non-purchase to purchase for top sponsored result and purchase to non-purchase for top organic results). 
This concern is even more critical since prior surveys show that almost half of the consumers are not able to distinguish between organic and sponsored results on Amazon~\cite{Graham2019Amazon}. Replacing top organic results with sponsored ones may be argued as deceptive design as they intend to exploit the cognitive and positional biases that consumers may have while interacting with digital mediums.

\vspace{1 mm}
\noindent
\textbf{Our position: }
Note that we do \textit{not} argue that the practice of advertising in general to be unfair. However, because of the competitive nature of the ads on digital marketplaces
the inadvertent consequences of interleaving advertisements with organic results need to be carefully thought through.
\section{Concluding discussion}

\textbf{Summary: }To our knowledge, this is the first work that tries to understand the effects and implications of sponsored results on the quality of search results in digital marketplaces. Our analyses of top sponsored results on Amazon digital marketplace across four large digital marketplaces show that they are not only low in relevance to the queries (as per Amazon's own ranking algorithm), but in most cases they are costlier and lower quality products than the corresponding organic results and thus degrading the overall result quality.

\noindent
\textbf{Amazon's consumer-centric approach:} While the analyses paints a bad picture of Amazon's sponsored product advertising policies, we must also acknowledge the different guardrails that Amazon has put in place to ensure consumer satisfaction. As per Amazon's sponsored product advertising policy, if at any time a seller is not winning a product's buy-box, her advertisements will not show up on Amazon SERP to consumers. 
At the same time, Amazon also ensures very high standards for selection of buy-box winners. However, while buy-box competition is among competing sellers for a single product, search ranking is a competition among different sellers through different products. Hence, mere restriction to buy-box winners may not lead to advertisements being beneficial to consumers.

\noindent
\textbf{Extending consumer-centric standards for selection and provision of ads: } Here are a few potential design choices which can serve all important stakeholders fairly:
\begin{enumerate}
    \item \textbf{Separation of sponsored and organic results:} One way to resolve the issue and adhere to different commercial laws is to separate out (i.e., stop complete interleaving of) sponsored results from organic results. 

    \item \textbf{Downgrade sponsored results:} Another relaxed version of the previous solution may be to slightly downgrade the sponsored results. For example, instead of replacing the top organic results, the sponsored results can come after the top few organic results. 

    \item \textbf{Guard rails of relevance and other features:} Along with some of these approaches, digital marketplaces should also consider stricter relevance criteria to select sponsored results, to prevent the deterioration of quality of search results. This may be achieved by using some constraints as to how far a sponsored result can be promoted from its organic position; its price and quality, etc. 
\end{enumerate}

\noindent
Adhering to such policies may resolve some of the unfairness concerns of existing advertising practices on competitive digital marketplaces, as well as bring in more competition and adherence to different fair market regulations.

\section*{Acknowledgements}
This research is supported in part by a European Research Council (ERC) Advanced Grant for the project ``Foundations for Fair Social Computing", funded under the European Union's Horizon 2020 Framework Programme (grant agreement no. 789373), and by a grant from the Max Planck Society through a Max Planck Partner Group at Indian Institute of Technology Kharagpur.

%\newpage
\bibliography{Main}

\end{document}